\documentclass[aps,twocolumn]{revtex4}

\usepackage{amsmath,amssymb}
\usepackage[bookmarks=false,colorlinks=true]{hyperref}
\usepackage{latexsym,bbm}
\newcommand{\sfrac}[2]{{\textstyle\frac{#1}{#2}}}
\newcommand{\ba}{\begin{array}}
\newcommand{\ea}{\end{array}}
\newcommand{\be}{\begin{equation}}
\newcommand{\ee}{\end{equation}}
\newcommand{\bea}{\begin{eqnarray}}
\newcommand{\eea}{\end{eqnarray}}
\newcommand{\Hzero}{\mathcal{H}_0}
\newcommand{\Hgenzero}{\mathcal{H}^\text{gen}_0}
\newcommand{\Hosc}{\mathcal{H}_{\omega}}
\newcommand{\Hcoul}{\mathcal{H}_\gamma}
\newcommand{\Hgenosc}{{\mathcal{H}}^\text{gen}_\omega}
\newcommand{\Hgencoul}{{\mathcal{H}}^\text{gen}_\gamma}

\begin{document}

\title{ Runge-Lenz vector in  the Calogero-Coulomb  problem}

\author{Tigran Hakobyan}
\email{tigran.hakobyan@ysu.am}
\author{Armen Nersessian}
\email{arnerses@ysu.am}
\affiliation{Yerevan State University, 1 Alex Manoogian St., Yerevan, 0025, Armenia}
\affiliation{Tomsk Polytechnic University, Lenin Ave. 30, 634050 Tomsk, Russia}

\begin{abstract}

 We construct the Runge-Lenz vector  and the symmetry algebra  of the rational
Calogero-Coulomb problem using the  Dunkl operators.
 We reveal that they are   proper deformations of their Coulomb counterpart.
 Together with similar correspondence between the Calogero-oscillator and oscillator models,
 this observation permits the  claim that most of the properties of the Coulomb  and oscillator systems
 can be lifted to their Calogero-extended analogs by the
 proper replacement of the momenta by the Dunkl momenta operators.
\end{abstract}

\maketitle

\section{Introduction}
The $N$-dimensional oscillator and Coulomb problems are the best known bound systems  with maximal
 number ($2N-1$) of functionally independent constants of motion. Such systems  are called maximally
superintegrable.
The free particle is the most widely known superintegrable unbound system.
It seems, that all other  superintegrable systems can be obtained somehow  from those listed above.

 The rational Calogero model \cite{calogero0,moser}
\be
\Hzero=
 \sum_{i=1}^N \frac{p_i^2}{2} + \sum_{i<j} \frac{g(g-1)}{(x_i-x_j)^2}.
\label{calo}
\ee
is highlighted  among  the non-trivial unbound superintegrable systems.
Its superintegrability
was  established  in the classical  \cite{woj83} and quantum \cite{kuznetsov,gonera}
cases.

The generalization of $\Hzero$, associated
with an arbitrary finite Coxeter group
\cite{rev-olsh}, is also superintegrable.Let us mention that the Coxeter group is described
 as a finite group  generated
 by a set of orthogonal reflections  across the hyperplanes $\alpha\cdot x=0$
 in the $N$-dimensional Euclidean space:
\be
\label{s-alpha}
s_\alpha x= x- \frac{2(\alpha\cdot x)}{\alpha\cdot\alpha}\alpha,
\qquad
\alpha\in \mathcal{R}_+\;.
\ee
Here the vectors $\alpha$ from the set $\mathcal{R}_+$  (called the system of positive roots)
uniquely characterize the reflections.
In this case the potential in \eqref{calo} is replaced by
\be
\label{coxeter}
V (x_1,\dots,x_N)= \sum_{\alpha\in\mathcal{R}_+}
\frac{g_\alpha (g_\alpha-1)(\alpha\cdot\alpha)}{2(\alpha\cdot x)^2}.
\ee
The coupling constants $g_\alpha$ form a reflection-invariant discrete function.
The original Calogero potential in \eqref{calo} corresponds to the $A_{N-1}$ Coxeter system with
the positive roots, defined in terms of the standard basis by $\alpha_{ij}=e_i-e_j$
for $i<j$. The reflections \eqref{s-alpha} become the coordinate
permutations in this particular case [see \eqref{sij} below].

The rational Calogero model with additional oscillator potential is  given by the Hamiltonian
\be
\label{osc}
\Hosc  =-\frac12\partial^2+\frac{\omega^2}2 x^2 +
 \sum_{i<j} \frac{g(g-1)}{(x_i-x_j)^2}.
\ee
 We refer to it as the \emph{Calogero-oscillator} model
\footnote{
Actually,  in the literature this system is referred to as the  Calogero
model, while the unbound system \eqref{calo} is referred as
the Calogero-Moser system due to Moser
who established the Liouville integrability \cite{moser}.
Our notations are more proper for reflecting the structure of underlying models.},
which is also superintegrable \cite{gonera-kosinski}.

The similarity between  the Calogero model  and a free particle,
as well as between the Calogero-oscillator model  and an oscillator,
 is clearly elucidated from the perspective of the matrix model reduction and
 the exchange operator formalism  (see \cite{rev-poly,rev-olsh} for the review).
Let us briefly outline the second approach, elaborated independently by
 Polychronakos \cite{poly92} and by Brink, Hansson,  and Vasiliev \cite{brink},
 which then has been found to be related with seminal work by Dunkl \cite{dunkl}.
Following these authors, we can take into account the Calogero interaction,
replacing the  momenta $ p_i=-\imath\partial_i $
by  the Dunkl momenta $-\imath\nabla_i$, defined in terms of the Dunkl
 operators \cite{dunkl}:
\be
\label{nabla}
\nabla_i=\partial_i-\sum_{j\neq i}\frac{g}{x_i-x_j}s_{ij} .
\ee
Here  $s_{ij}$ is the exchange operator between the $i$th and $j$th coordinates:
\be
\label{sij}
s_{ij}\psi(\dots, x_i,\dots, x_j, \dots) =\psi(\dots, x_j,\dots, x_i, \dots).
\ee
Amazingly, such operators commute like usual partial derivatives:
\be
\label{com-nn}
\qquad
[\nabla_i,\nabla_j]=0.
\ee
Meanwhile, their commutations with the coordinates are more involved
and expressed through the permutations:
\be
\label{com-nx}
[\nabla_i, x_j]=S_{ij}=
\begin{cases}
 -g s_{ij}&\text{for $ i\ne j$},\\
1+g\sum_{k\ne i} s_{ik} & \text{for $i= j$}.
\end{cases}
\ee
In the absence of the inverse-square potential ($g=0$), the
above relations define the usual Heisenberg algebra.

The Calogero-oscillator Hamiltonian \eqref{osc} can be obtained by the
restriction of the generalized Hamiltonian
\be
\label{gosc}
\Hgenosc=-\frac{1}{2}\nabla^2+\frac{\omega^2}{2}   x^2
\ee
 to the  symmetric wavefunctions \cite{brink}:
 \be
 \label{sym}
 s_{ij}\psi_s(x)=\psi_s(x)
 \ee
In these terms  there is a remarkable similarity between the
integrals of motion of the Calogero-oscillator and ordinary oscillator systems.
First, take the  overcompleted set of the symmetry generators of $\Hgenosc$,
which are given by the  Dunkl angular momentum  operator \cite{kuznetsov,feigin}
\footnote{For simplicity, we omit the imaginary unit $\imath$ in the definition  \eqref{Mij}
so that $M_{ij}$ becomes anti-Hermitian in our notations.
}
\be
\label{Mij}
 M_{ij}=x_i\nabla_{j}-x_j\nabla_{i},
\ee
satisfying the deformed commutation relations  \cite{kuznetsov,fh}
\be
\label{Mcom}
[M_{ij},M_{kl}]=S_{kj}M_{il} + S_{li}M_{jk}
- S_{ki}M_{jl} - S_{lj}M_{ik}  ,
\ee
and the hidden symmetry generators
\be
\label{Iij}
I_{ij}=-\nabla_i\nabla_j+ \omega^2 x_ix_j .
\ee
Note that in  the $g=0$ limit, the generators $M_{ij}$ and $I_{ij}$ are reduced to the
unitary  algebra $u(N)$,
which  describes the symmetry of the
$N$-dimensional isotropic oscillator.
The constants  of motion of the  Calogero-oscillator model $\Hosc$
can be associated with the symmetric polynomials
\begin{gather}
\label{Mcal}
\mathcal{M}_{2k}=\sum_{i<j}M_{ij}^{2k},
\\
{\cal I}_{k}=\sum_{i}^N I_{ii}^k,
\qquad
{\cal I}'_{k}=\sum_{i<j}^N I_{ij}^k.
\end{gather}
The symmetrization ensures their valid action on the  wavefunctions
\eqref{sym}.

The symmetries of  the Calogero model without oscillator $\Hzero$ are related to
the symmetries of the free-particle system in the same way.
Being restricted to the symmetric wavefunctions, the Hamiltonian \eqref{calo} coincides with the
analog of the free-particle Hamiltonian with the Dunkl derivatives used instead
of the standard ones \cite{poly92}:
\be
\label{gcal}
\Hgenzero=-\frac{1}{2}\nabla^2.
\ee
It commutes both  with the Dunk operators \eqref{nabla} and
the Dunkl momenta generators \eqref{Mij}. The symmetric polynomials
\be
\label{liouville}
I_k|_{\omega=0}=\sum_i \nabla_i^k
\ee
mutually commute, in contrast to
 the Dunkl angular momentum polynomials \eqref{Mcal}.
For $k\le N$ they form the set of the Liouville integrals of motion of the Calogero Hamiltonian \eqref{calo}.
The  operators \eqref{Mcal} complete them
to the  full set of integrals of motion \cite{poly92}.

Hence, the symmetries of the rational Calogero model without and with the oscillator potential, formulated in terms
of the Dunkl operators,  are in one-to-one correspondence with those  of the  free particle and
 the oscillator,  respectively.
This holds  also for a more general rational potential \eqref{coxeter},
associated with an arbitrary Coxeter group \cite{fh}.

On the other hand, it has been known for  years that the rational Calogero model,
extended  by any other central potential, remains an integrable system  \cite{khare,khare99}.
The integrability is more or less obvious. It  proceeds from
the integrability of the angular  part of the generalized rational Calogero model \cite{saghatel}.
Meanwhile, only few of them preserve the superintegrability property.
In a recent paper with Lechtenfeld,  we have shown
that the oscillator and Coulomb potentials are unique in this context \cite{CalCoul}.
%
 Moreover, we have observed that the Calogero-oscillator and \emph{Calogero-Coulomb}
 models are, in fact, the only isospectral deformations of the
 original oscillator and Coulomb  systems.

Let us remember that the hidden symmetries of the Coulomb problem are given by the Runge-Lenz vector,
forming a quadratic algebra together with the angular momentum operators \cite{revai65}.
It is reduced to the orthogonal $so(N+1)$ or pseudo-orthogonal $so(N,1)$ algebra,
respectively, on the bound (negative-energy) or unbound (positive-energy) states.

Having in mind the similarity between the symmetries  of the isotropic oscillator and
Calogero-oscillator models, and the fact that  the Calogero-oscillator
and Calogero-Coulomb systems  are highlighted from an integrability viewpoint  \cite{CalCoul},
 we can ask
\emph{whether the symmetries of the conventional Coulomb problem  can be deformed
to the symmetries of   the Calogero-Coulomb model.}

\medskip

In this article we will show the following:
 \begin{enumerate}
\item The symmetry generators of the Calogero-Coulomb system, formulated in terms of
 the Dunkl operators,  are given  by the deformed angular momentum tensor \eqref{Mij}
 and by the  deformed Runge-Lenz vector.
\item
The symmetry algebra of the Calogero-Coulomb model is a deformation of
the symmetry algebra of the initial Coulomb problem.
\item
The functional relation between the Coulomb Hamiltonian,  Runge-Lenz vector and
the angular momentum has  an analog in the Calogero-Coulomb problem.
\end{enumerate}

In fact, this means that  the Calogero-Coulomb problem is as fundamental as
the Calogero-oscillator problem.
Due to such profound similarity with the conventional Coulomb problem,
we expect that most of the applications of the Coulomb system
can be extended somehow to the Calogero-Coulomb system.

\section{Symmetries }
In this section we demonstrate that all symmetries of the $N$-dimensional
quantum Coulomb model
can be deformed to those of the Calogero-Coulomb problem
\cite{khare,khare99}
\footnote{In our notations, the subscript in $\Hosc$ and $\Hcoul$ is not just an argument, but it also defines
the type of  confining potential.},
\be
\label{coul}
\Hcoul  =-\frac12\partial^2 -   \frac{\gamma}{r} +
 \sum_{i<j} \frac{g(g-1)}{(x_i-x_j)^2},
\ee
using the
Dunkl operator formalism.

The generalized Calogero-Coulomb model  is described by the following Hamiltonian:
\be
\label{gcoul}
\Hgencoul=
-\frac{\nabla^2}{2}- \frac{\gamma}{r}
\qquad
\text{with}
\qquad
r= \sqrt{x^2}.
\ee
As  in the previously discussed Calogero-oscillator case,
it preserves the Dunkl angular momentum operators:
\be
[\Hgencoul,M_{ij}]=0.
\ee
We define the following deformation of the ordinary Runge-Lenz vector:
\be
\label{Ai}
 A_i=-\frac12 \sum_j\left\{ M_{ij},\nabla_j\right\}
+\frac12[\nabla_i,S]
 -\gamma\frac{x_i}{r},
 \ee
 where
 \be
 \label{S}
  S=\sum_{i<j}S_{ij}.
\ee
Here and in the following, the curly brackets mean an anticommutator
of two operators:
\be
\{a,b\}=ab+ba.
\ee
The operator $S$ is invariant with respect to the permutations
and is a constant on the symmetric wavefunctions \eqref{sym}:
\be
 [S,s_{ij}]=0,
 \qquad
 S\psi_s(x)=g\frac{N(1-N)}{2}\psi_s(x).
 \ee
In the absence of inverse-square interaction, $g=0$, the operators $S_{ij}$ are reduced to
$\delta_{ij}$, and the second term in \eqref{Ai} vanishes. Respectively, the conserving  quantities
$A_i$  are  reduced to the usual Runge-Lenz vector of the $N$-dimensional Coulomb system.

In the Appendix we prove that the generalized Runge-Lenz vector provides the system
with the hidden symmetry:
\be
\label{comAH}
[\Hgencoul, A_i]=0.
\ee
Therefore, the operators $M_{ij}$ and $A_i$ generate entire
symmetry algebra of the generalized Calogero-Coulomb system
\eqref{gcoul}.
It appears that they obey the following commutation relations:
\be
\label{comAM}
\begin{split}
&[A_i,M_{kl}]=A_{l}S_{ki}-A_{k}S_{li},
\\
&[A_i,A_j]=-2\Hgencoul M_{ij}.
\end{split}
\ee
At the pure Coulomb point  (i.e. in the $g=0$ limit)  these relations together with \eqref{Mcom}
are reduced  to the  symmetry algebra of the $N$-dimensional Coulomb problem.

The second commutation relation in Eq.~\eqref{comAM} is proven in the Appendix.

Consider the first commutator in \eqref{comAM}. It can be viewed
as a deformation
of the infinitesimal rotation of the vector  $A_i$.
Note that the  coordinates ($u_i=x_i$) and
Dunkl operators  ($u_i=\nabla_i$) obey the same relation, as it follows from
Eqs. \eqref{nabla}, \eqref{com-nn}, \eqref{com-nx},
 and \eqref{Mij}:
\be
\label{vector}
[u_i,M_{kl}]=S_{ik}u_{l}-S_{il}u_{k}
=u_{l}S_{ki}-u_{k}S_{li}.
\ee
Now we express $A_i$ in terms of the
coordinates and Dunkl operators. The following formula is proven in the Appendix:
\be
\label{Ai-2}
A_i=\Big(r\partial_r+\frac {N-1}2\Big)\nabla_i - x_i \Big(\nabla^2+\frac\gamma r\Big).
\ee
Evidently, the operator-valued coefficients of $x_i$ and $\nabla_i$ in the
above expressions
commute with $M_{kl}$ and $S_{kl}$. Hence, the first commutation relation in \eqref{comAM}
follows directly from the identities \eqref{vector}.

Like in  the oscillator case, we are forced to combine the conserving quantities
$A_i$ and $M_{ij}$
 into the symmetric polynomials
\be
\label{Ak}
{\cal A}_k=\sum_{i=1}^N A_i^k.
\ee
and \eqref{Mcal} in order to get the well-defined constants of motion for the original
model \eqref{coul}.

The first member of the family set \eqref{Ak}
is independent of  the $S$-term  and is given by the expression
\cite{CalCoul}
\begin{gather}
\label{A1}
{\cal A}_1 =  \sum_i x_i\,\Big(2\Hgencoul +\frac\gamma r\Big)
+ \Big(r\partial_r + \frac{N-1}{2}\Big)\sum_i\partial_i.
\end{gather}
The constant of motion $\mathcal{M}_2$   does not commute
with $M_{ij}$ but  is related  with  the  Casimir element $\mathcal{M}'_2$
of the algebra \eqref{Mij}
in a rather simple way \cite{fh}:
\be
\label{Msq}
\begin{split}
\mathcal{M}'_2&=
 \mathcal{M}_2 -S(S-N+2)
 \\
& = r^2\nabla^2 -  r^2\partial_r^2 -(N-1)r\partial_r.
\end{split}\ee
It describes the angular part of the Calogero model,
studied  from various perspectives in \cite{sphCal,flp}.

The constant of motion $\mathcal{A}_2$  does not commute with $M_{ij}$ as well.
However, the corrected integral
\be
\label{A2'}
\mathcal{A}'_2=\mathcal{A}_2+2\Hgencoul S
\ee
 becomes commutative with  the Dunkl angular momentum, as was
proven in the Appendix:
\be
[\mathcal{A}'_2,  M_{ij}]=0.
\ee
This suggests that a certain combination of these  invariants may commute with  $A_i$ too.

In the Appendix we prove the following relation between the symmetry generators,
 which generalizes a similar relation in the conventional Coulomb problem:
\be
\label{cas}
\mathcal{A}'_2
=\gamma^2-2\Hgencoul\big( \mathcal{M}'_2-\sfrac{(N-1)^2}{4}\big).
\ee
Presumably, it can be used for the pure algebraic derivation of the spectrum of the Calogero-Coulomb problem.

\section{Concluding remarks}
In this article we have proven that all relations between the symmetry generators of the Coulomb problem
can be extended to the Calogero-Coulomb model.
To obtain them we should  replace the momenta operators $-\imath\partial_i$
by the Dunkl momenta $-\imath\nabla_i$
and make proper corrections depending on the permutation operators.
It is straightforward to extend our consideration to  the Calogero-Coulomb model
associated with arbitrary Coxeter group. Note that the two-dimensional Calogero-Coulomb
problem associated with the dihedral group $D_2$ was investigated recently using the Dunkl operators
\cite{vinet}.
The same correspondence holds  for the Calogero-oscillator  model \cite{fh}.

Both the Calogero-oscillator and Calogero-Coulomb models have superintegrable counterparts on
 (pseudo)spheres \cite{CalCoul},
and we have no doubt that the symmetry algebras of (pseudo)spherical oscillator and Coulomb systems can be
 lifted, in the same way, to those with Calogero term.
However, due to technical difficulties we  are  unable to complete these calculations.

This remarkable similarity between the Calogero-oscillator (Calogero-Coulomb) model and
oscillator (Coulomb) permits us  to claim that
\emph{most of the properties of the oscillator and Coulomb systems can
be extended to their counterparts supplemented by the Calogero interaction term.}
We are sure that in this way one can  construct  the superintegrable extensions of three- or five-dimensional
Calogero-Coulomb problems, specified by the presence, respectively, of the Dirac and Yang monopoles.
Moreover, it seems that acting in the  suggested way, we can relate
the two-, four-, and eight-dimensional
Calogero-oscillator models with the two-, three-, and five-dimensional Calogero-Coulomb models,
including those specified by the presence of anyon and  Dirac, Yang monopoles in the spirit
of Ref.~\cite{pogosian}. (For previous treatments see Ref.~\cite{khare99}.)

Recently,  the superintegrability of the (relativistic) Dirac oscillator \cite{Do}
and  Coulomb \cite{Dh} systems has been established.
It would be interesting to study    them in the presence of
Calogero interaction from the superintegrability point of view.
Moreover,  we expect that in  this way  one can solve the problem of ${\cal N}=4,8$
supersymmetrization of the Calogero model, which was treated  by many authors
(see \cite{susy} and references therein).

\acknowledgments
This work was partially supported
by the Armenian State Committee of Science Grant No. 15RF-039.
It was completed within programs of the Regional Training Network on Theoretical Physics
sponsored by Volkswagenstiftung under Contract No. 86 260 and of the
ICTP Network NET68.

\appendix

 \section{Derivations }

\subsection{Conservation of Runge-Lenz vector \eqref{Ai} }
Here we prove that the Calogero-Coulomb Hamiltonian preserves the deformed Runge-Lenz vector
\eqref{Ai} .
First we  compute the commutator  between the Hamiltonian  and  $ A_i$.
The commutator with the first term in  the right hand side of the equation \eqref{Ai}
can be  simplified using the following identity:
\begin{multline}
\Big[\sum_{j}\{M_{ij},\nabla_j\}\,,\,\frac{1}{r} \,\Big]
= - \frac{1}{r^3}  \sum_j \left\{M_{ij},x_j\right\}
 \\
 =   \Big\{\frac{1}{r}, \nabla_i\Big\}- \sum_j \left\{ \frac{x_ix_j}{r^3}, \nabla_j\right\}.
\label{AHpr1}
\end{multline}
In the derivation we have used the vanishing of the two commutators \cite{fh}:
\be
[\nabla^2,M_{ij}]=[r,M_{ij}]=0.
\ee

Next,  we can calculate the commutator of
the Hamiltonian with the last term in the deformed Rune-Lenz vector
expression   \eqref{A1} using the following identity:
  \begin{gather}
 \label{AHpr2}
 \begin{split}
\Big[\frac{x_i}{r} , \nabla^2 \,\Big]
 =    \sum_j\Big\{ \frac{x_ix_j}{r^3}-\frac{S_{ij}}{r},\nabla_j\Big\}.
 \end{split}
  \end{gather}
Combining together the relations \eqref{AHpr1} and \eqref{AHpr2},
we obtain:
\begin{multline}
\label{final}
\Big[ -  \frac12\sum_{j}  \{M_{ij},   \nabla_j\}-\frac{\gamma x_i}{r}\,, \, \Hgencoul\Big]
\\
\quad =   \sum_{j}\left\{\frac{\gamma S_{ij}}{2r}, \nabla_i-\nabla_j\right\}
=\frac12\big[ [S,\nabla_i],\Hgencoul\big],
\end{multline}
where $S$ is defined in \eqref{S}.
In the last equation we have used the identity
\[
\sum_j(\nabla_j-\nabla_i)S_{ij}=[S,\nabla_i].
\]
The relation \eqref{final} completes the proof of conservation  of the deformed
Runge-Lenz operator \eqref{comAH}.

\subsection{Second commutation relation in  \eqref{comAM}}
Let us derive the  commutation relation between the components
of the  Runge-Lentz vector in  \eqref{comAM}.
For convenience, we present the deformed Runge-Lenz vector
\eqref{Ai-2} in the following form:
\be
\label{Aab}
A_i=a\nabla_i-x_ib,
\quad
a\equiv r\partial_r+\frac{N-1}{2},
\quad
b\equiv \nabla^2+\frac{\gamma}{r}.
\ee
Then
\be
\begin{split}
\label{AiAj}
[A_i,A_j]=&[a\nabla_i,a\nabla_j]+[x_ib,x_jb]+[a\nabla_{j},x_{i}b]
\\
&-[a\nabla_{i},x_{j}b].
\end{split}
\ee
Note that the  commutator $[a,f]$ counts the total degree in coordinates of
the quantity $f$:
\be
\label{com-a}
[a,\nabla_i]=-\nabla_i,
\quad
[a,x_i]=x_i,
\quad
[a,b]=-2\nabla^2-\frac\gamma r.
\ee
The commutators in \eqref{AiAj}, containing the observable $b$, are
simplified to
\be
\label{com-b}
[b,x_i]=2\nabla_i,
\qquad
[b,\nabla_i]=\frac{\gamma x_i}{r^3}.
\ee
Using the above equations, we obtain:
\begin{gather}
\label{same}
[a\nabla_i,a\nabla_j]=0,
\qquad
[x_ib,x_jb]=2M_{ij}b,
\\
\label{mixed}
[a\nabla_j,x_ib]
=x_j\nabla_i\nabla^2+\frac{\gamma  x_ix_j}{r^3}(a+1)+aS_{ij}b.
\end{gather}
The last two terms on the right hand side of \eqref{mixed} are symmetric on
the indexes $i$ and $j$, and, hence,  disappear
in the commutator \eqref{AiAj}.
Substituting Eqs. \eqref{same} and  \eqref{mixed} into \eqref{AiAj}, we arrive at the relation
sought:
\be
[A_i,A_j]=-2\Hgencoul M_{ij}.
\ee

\subsection{Relation \eqref{Ai-2} }
Here we derive the relation \eqref{Ai-2}, which expresses the Runge-Lenz invariant  in terms of the
coordinates and Dunkl momenta.
First we calculate the first term of \eqref{Ai}:
\begin{multline}
\sum_j\{ M_{ij},\nabla_j\} =\{\nabla^2,x_i\}-(x\cdot\nabla)\nabla_i- \nabla_i(\nabla\cdot x)
\\
=\{\nabla^2,x_i\}-\big(2r\partial_r+(N+1)\big)\nabla_i+[\nabla_i,S].
\end{multline}
We have used the following identities in the derivation:
\begin{gather}
\label{x-nabla}
x\cdot\nabla  = r\partial_r + S,
\qquad
\nabla \cdot x= r\partial_r - S+N.
\end{gather}
Finally, substituting them into Eq. \eqref{Ai}, we arrive at the equation \eqref{Ai-2}.

\subsection{Relation  \eqref{A2'} }
Let us calculate the commutator of $M_{ij}$ with $\mathcal{A}_2$:
\be
\label{M-Asq}
[M_{ij},\mathcal{A}_2]
=\sum_k \{A_{i}S_{jk}-A_j S_{ik} , A_k\}.
\ee
Each term from the right-hand side  of this equation can be presented as
\begin{gather}
\begin{split}
\sum_k\{A_iS_{jk},A_k\}
=2A_iA_j-2\Hgencoul \sum_k M_{ki}S_{kj}.
\end{split}\end{gather}
Here we take into account the identity
\be
\sum_k\{S_{ik},u_k\}=2u_i+\sum_{k\ne i}\{ S_{ik},u_k-u_i\}=2u_i,
\ee
which is fulfilled for any local operator $u_k$.
Applying  the above relation, one can further simplify the commutator \eqref{M-Asq}:
\begin{align}
[M_{ij},\mathcal{A}_2]&=
-2\Hgencoul\Big (\sum_k( M_{ki}S_{jk} - M_{kj}S_{ik})
+ 2M_{ij}\Big)
\nonumber\\
&=-2\Hgencoul \Big[M_{ij},\sum_{k\ne i, j}(S_{kj}+S_{ki})\Big]
\nonumber\\
&=-2\Hgencoul [M_{ij},S].
\end{align}
This completes the proof of the equation \eqref{A2'}.

\subsection{Relation \eqref{cas}}
We use the representation \eqref{Aab} for $\mathcal{A}_2$:
\be
\label{A2ab}
\begin{split}
\mathcal{A}_2=&\sum_i (a\nabla_i - x_i  b)^2
 =  (a+1)a \nabla^2
 \\
 &+r^2b^2+2(x\cdot\nabla) b  - \sum_i \{a \nabla_i,x_ib\}
\\
=&r^2b^2+ 2a^2\Hgencoul-2(x\cdot\nabla) \mathcal{\Hgencoul}-2a\frac{\gamma}{r}.
\end{split}
\ee
The commutation relations \eqref{com-a}, \eqref{com-b} and \eqref{x-nabla} are used in the derivation.

In the first term in the last expression one can select the Hamiltonian as follows:
\be
r^2b^2
=-2r^2\nabla^2\Hgencoul+\frac{\gamma(N-3)}{r}+2\gamma\partial_r+\gamma^2.
\ee
Inserting this into Eq. \eqref{A2ab} and simplifying it,  we arrive at the desired
relation \eqref{cas}.

\end{document}